\documentclass[aps,twocolumn]{revtex4}
\usepackage{epsfig}
\newcommand{\be}{\begin{equation}}
\newcommand{\ee}{\end{equation}}
\newcommand{\bea}{\begin{eqnarray}}
\newcommand{\eea}{\end{eqnarray}}

\begin{document}

\title{Extreme-value statistics and super-universality in critical percolation?}

\author{Mohadeseh Feshanjerdi}\email{m.feshanjerdi@alzahra.ac.ir}
\affiliation{Department of Condensed Matter Physics, Faculty of Physics, Alzahra University, P. O. Box 1993893973, Tehran, Iran}
\author{Peter Grassberger}\email{p.grassberger@fz-juelich.de}
\affiliation{JSC, FZ J$\ddot{u}$lich, D-52425 J$\ddot{u}$lich, Germany}
\affiliation{Max Planck Institute for the Physics of Complex Systems, N\"othnitzer Str.38, 01187 Dresden, Germany}
\date{\today}

\begin{abstract}
	Recently, the number of non-standard percolation models has proliferated. In all these models,
	there exists a phase transition at which long range connectivity is established, if local 
	connectedness increases through a threshold $p_c$. In ordinary (site or bond)
	percolation on regular lattices, this is a well understood second-order phase transition with
	rather precisely known critical exponents, but there are non-standard models where the transitions
	are in different universality classes (i.e. with different exponents and scaling functions), 
	or even are discontinuous or hybrid. It was recently claimed that certain scaling functions are 
	in all such models given by extreme-value theory and thus independent of the precise universality
	class. This would lead to super-universality (even encompassing first-order transitions!) and 
	would be a major break-through in the theory of phase transitions. We show that this claim
	is wrong.
\end{abstract}

\maketitle

\section{Introduction}

Percolation \cite{Stauffer,Sahimi} is one of the paradigmatic topics in statistical physics.
It is concerned with the appearance or breakdown of long-range connectivity, as local connections
are established or removed. This can be the formation of a gel if bonds between molecules are 
increased, the spreading of a disease if contact between people is increased above a threshold, 
or even the acceptance of a scientific theory as the number of its supporters surpasses a critical
level. It is thus not surprising that percolation is a topic discussed in all branches of science,
from condensed matter physics to astrophysics, biology, finance, and sociology \cite{Saberi,Araujo}.

What makes percolation an interesting object of research is that the transition 
to long-range connectivity is, in the large system limit, abrupt with all features usually
attached to a phase transition. In what we shall call in the following ``ordinary percolation" (OP),
this is a well-understood continuous (`second order') transition, whose critical exponents are 
exactly known in $d=2$ and for $d>6$, and known to high precision in dimensions three to five.
The best known examples of OP are bond and site percolation on regular lattices \cite{Stauffer,Sahimi}. 

Just like other critical phenomena, also OP enjoys what is known as universality. This means that
properties like critical exponents and scaling functions are strictly independent of details like
the type (but not the dimension) of the lattice, and whether local connections are broken by 
removing sites or bonds. A set of microscopic models which thus have identical scaling behavior 
is called a universality class.

But the very wide conceptual range of the notion of percolation implies that there are systems
which fall outside the standard universality classes or, otherwise said, there are different 
universality classes. Examples are rigidity \cite{rigidity}, bootstrap \cite{Adler,di_Muro},
correlated \cite{Weinrib,Schrenk} (e.g. pacman \cite{Abete} and aftermath \cite{aftermath}) and
explosive \cite{Achlioptas} percolation, percolation on random (Erd\"os-Renyi) networks \cite{Newman}, 
agglomerative percolation \cite{Lau}, percolation driven by random walks \cite{Feshanjerdi},
systems where the `local' bonds can be arbitrarily long and have power-law distributions 
\cite{Grassberger2012,Grassberger2013,Gori}, and systems where several local elements have to 
cooperate in order to establish connections \cite{Dodds,Dorogovtsev2006,Bizhani,Cai}. In some 
of these cases \cite{Adler,di_Muro,Dodds,Dorogovtsev2006,Bizhani,Cai} 
the differences to OP are so big that the transition is discontinuous (`first order').

Even when two systems are in different universality classes, it may happen that for some reason
one or more critical exponents (or scaling functions) are identical, which is then called 
{\it super-universality}. The best known examples are provided by surface critical phenomena 
\cite{Cardy}. Take e.g. site percolation on a semi-infinite square lattice with $-\infty < x,y < \infty,
0 < z < \infty$. In this case clusters growing from a surface site with $z=0$ have different 
fractal dimensions than clusters deep in the bulk, but correlations are described by the usual 
(bulk) correlation length exponent $\nu$, even very close to the surface \cite{surface-percol}. 
The same happens for the Ising model and for self-avoiding random walks. Another example is 
random interfaces within the so-called Kardar-Parisi-Zhang \cite{KPZ} universality class. Here 
the critical exponents are independent of the genus of the surface (asymptotically flat, circular, ...), 
but scaling functions do in general depend on the genus \cite{Takeuchi}. Finally, we might mention 
polymer networks \cite{Duplantier} where the Flory index controlling the size is independent of 
the topology, but the exponents controlling the free energy are not.

Recently, it was claimed \cite{Fan} that such super-universality is for percolation much more common 
than previously thought. The reason, it is claimed in \cite{Fan}, is the following. Usually, a critical
point $p_c$ is defined as the point in control parameter space where observables or partition sums of 
{\it infinite} systems would develop singularities. One can define effective transition points also 
for finite systems, but in the standard treatment such definitions are always ad hoc and heuristic.  
For percolation problems, a more principled way to define transition points for finite systems 
consists in looking at jumps in the mass of the largest cluster, as local connections are added one 
by one \cite{Manna} (see also \cite{Nagler,Lee}). If the system is far subcritical, these jumps are 
small, since clusters are anyhow small. They are also small in far supercritical systems, since there the 
giant cluster can grow only by `eating' much smaller ones. But near the critical point, giant clusters 
can grow by merging two or more large clusters, and thus jumps will be large. Thus one can define a 
transition point $p_{\rm max}({\cal E})$ for 
each (finite) event ${\cal E}$ as the point where the largest jump (called the `gap' in the 
following) occurs in ${\cal E}$. Here, an `event' ${\cal E}$ is the entire evolution of one
realization of the system.

In standard finite size scaling (FSS) analyses, observables are considered as functions of 
$p-p_c$. In addition, one can study the distributions of observables at $p=p_c$, like the masses
of the largest and second-largest clusters. 
The claim made in \cite{Fan} is that several scaling functions become 
super-universal in a very wide class of non-standard percolation models, if FSS analysis is 
based on $p_{\rm max}({\cal E})$. To be more precise, consider the distribution of gap sizes 
over the entire ensemble. By definition, the gap size of ${\cal E}$ is the size of the 
largest jump in the entire history of the event, and $p_{\rm max}({\cal E})$ is its 
location in the history. Thus a gap is defined as an extremal value
in a set of jumps (all jumps in the given history). In the limit of large systems this set becomes
infinite, and the jumps in one history are basically unbounded and uncorrelated. Thus the gap
(the extremal value in this set) should be distributed according to a Gumbel (Fisher-Tippett type 1)
distribution \cite{Gumbel}. Since this should hold for all (generalized) percolation models,
the gap distribution should be super-universal. The weak point in this argument is, of course, 
that jumps are unbounded and uncorrelated only for infinite systems, while we are dealing with
large but finite systems.

It was indeed claimed in \cite{Fan} that not only the gap sizes obey a Gumbel distribution,
but also the masses of the largest cluster at $p_{\rm max}({\cal E})$. Finally, the transition 
points $p_{\rm max}({\cal E})$ themselves were claimed to obey a Gauss distribution.

The latter two claims were made without strong mathematical arguments, but also in case of 
the gap distribution the argument is very weak indeed. Since the sum of all jumps in an event 
is constrained to be the total mass of the system, it is clear that the jumps in an event 
cannot be uncorrelated. It might be a priori possible that these correlations are sufficiently 
weak that the distribution is still Gumbel, but this is by no means guaranteed. 

Indeed, Fan {\it et al.} were not the first ones to claim that the Gumbel distribution is relevant 
for percolation. Bazant \cite{Bazant} showed that the masses of the largest cluster in
{\it subcritical} OP follow a Gumbel distribution, in the infinite system limit. Although
not rigorously provable \cite{Bazant}, this makes perfect sense, since in this case the variables 
are effectively uncorrelated and unlimited. A similar but less convincing claim was made by 
Bramwell {\it et al.} \cite{Bramwell}, who considered the masses of the largest cluster in nearly 
critical OP. Precisely at the critical point, the distribution of the largest cluster is well
known to follow a power law for small masses, and to have a rather sharp peak at large 
masses \cite{Stauffer}. As one decreases $p$, the peak first becomes a shoulder, and then
the two merge to form something resembling superficially a Gumbel distribution. But this is 
far from an exact identity, unless one strictly follows Bazant's strategy of keeping a fixed 
$p < p_c$ and taking $N\to\infty$. While it is crucial for \cite{Bazant} that different regions 
in a large subcritical percolation problem are essentially uncorrelated, \cite{Bramwell} claim
that their finding holds in an ``inertial range", and stress that ``One of the
anomalous statistical properties of inertial systems is that, whatever their size, they can 
never be divided into mesoscopic regions that are statistically independent". They are right 
that this excludes the validity of Gauss' central limit theorem, but it also prevents the 
cluster size distributions from being Gumbels.

In the rest of this paper we shall investigate the claim made by Fan {\it et al.} numerically,
by studying six different universality classes. Of these six, five show continuous transitions, 
while one (the cluster aggregation (DCA) model of \cite{Cho}) has a discontinuous transition. For 
the latter, it is immediately seen that the claim of Fan {\it et al.} is wrong, and we thus 
show only data for two system sizes. Among the other five models, three are on lattices 
(2-d and 3-d site percolation and aftermath percolation in $d=3$ \cite{aftermath}), while two (
Erd\"os-Renyi (ER) and product-rule explosive percolation (PR\_EP)) are off-lattice. 
Simulation data for five of these universality classes (DCA,ER,PR\_EP, 2-d and 3-d OP) had been 
used by Fan {\it et al.} to support their claim. We find that in all cases (except possibly PR\_EP) 
the distributions of $p_c({\cal E})$ are wider than Gaussians, although the differences are not 
dramatical. Much more significant are the differences among the gap distributions. Except possibly 
for ER, they are in all cases significantly more narrow than Gumbels, and differ widely from model 
to model. The same is true, although to a smaller amount, for mass distributions. While the sizes of 
the systems simulated in the present paper are comparable to those in \cite{Fan} or even slightly 
smaller, it seems that our statistics are vastly superior (for the largest systems we made 
$>4\times 10^5$ simulations for each model, for the smallest ones $>4\times 10^6$).

\section{Simulations and Finite Size System Analysis}

All six models were simulated by means of the Newman-Ziff (NZ) algorithm \cite{Newman-Ziff}, because 
in this way we had the fastest and easiest access to jumps in the masses of the largest cluster.
In this algorithm, one starts with an `empty' system, and sites or bonds are inserted one after 
another. The mass of the largest cluster after the $n-$th insertion is $S_n$, and the $n-$th 
jump is defined as $\Delta_n = S_n-S_{n-1}$. For each model we denote the control parameter by 
$p = n/N$, where $N$ is the total number of sites and $n$ is the number of already inserted 
sites/bonds.  For the lattice models (2-d and 3-d site and aftermath) $p$ is the density of
inserted sites, while $p$ is the fraction of established bonds for the off-lattice models 
(ER, PR\_EP, and DCA). For the latter (DCA), a rather complicated algorithm was described in 
the supplementary material \cite{suppl}
of \cite{Fan}. We did not use this, but we used the simple algorithm described in \cite{Cho}. 
Nevertheless, our data seem to agree perfectly with the simulation data of  \cite{Fan}, even 
though our interpretation of these data differs radically.

For the lattice models, we used helical boundary conditions, i.e. sites were (irrespective of 
the dimension $d$) indexed by a single integer $i$, neighbors were $i\pm 1, i\pm L, \ldots 
i\pm L^{d-1}$, and $i+N \equiv i \;{\rm mod}\; N$ with $N=L^d$. System sizes were $L=512, 1024, 
2048$, and 4096 for 2-d site OP, $L=32,64,128$, and 256 for 3-d lattices, and $N=2^{16},2^{18},
2^{20}$, and $2^{22}$ for ER and PR\_EP. For DCA we show results only for $N=2^{16}$ and 
$N=2^{18}$, since the conclusions from this were obvious anyhow.

The gap $\Delta_{\rm max}$ is defined the largest jump during a run of the NZ algorithm.
Assume that it occurs at the $n-$th insertion. This defines the event-based transition point
as $p_{\rm max}({\cal E}) = n/N$, and $S_{\rm max}$ is the mass of the largest cluster after the 
$n-$th insertion. 

For each model (and for each system size) we show the three distributions 
$P_p(p_c), P_\Delta(\Delta_{\rm max})$, and $P_S(S_{\rm max})$. According to the standard FSS
scaling assumptions, they are supposed to satisfy for finite-dimension models the scaling laws
\be
   P_S(S_{\rm max}) = L^{-d_f} f_S(S_{\rm max}/L^{d_f}),       \label{P_S}
\ee
\be
   P_p(p_{\rm max}) = L^{1/\nu} f_p[(p_{\rm max}-p_c)L^{1/\nu}],  \label{P_p}
\ee
and
\be
   P_{_\Delta}(\Delta_{\rm max}) = L^{-d_f} f_{_\Delta}(\Delta_{\rm max}/L^{d_f}),  \label{P_Delta}
\ee
where $d_f$ is the fractal dimension, $\nu$ is the correlation length exponent, and the functions
$f_S, f_p$, and $f_{_\Delta}$ are universal scaling functions. For ER and PR\_EP, we use the same 
ansatzes but with $L$ replaced by $N$ and with $d_f$ replaced by the scaling exponent $\eta$ in the 
relationship $S_{\rm max} \sim N^\eta$, and $\nu$ must be defined by $p_{\rm max}-p_c \sim N^{-\nu}$.

Equations (1-3) correspond to Eqs.~(11-13) of \cite{Fan}, but some comments are in order:\\
(i) Eqs.~(1) and (3) follow immediately from the assumption that scaling functions $f_i$ must 
have dimensionless arguments and that $S_{\rm max}$ and $\Delta_{\rm max}$ both are masses.
Similarly, Eq.~(2) follows from the fact $(p_{\rm max}-p_c)L^{1/\nu}$ is essentially the unique 
dimensionless variable involving $p_{\rm max}$. The arguments put forward in \cite{Fan} seem 
more complicated.\\
(ii) In Eqs.~(11) and (13) of \cite{Fan}, the scaling functions are assumed to be functions of 
$S_{\rm max}-\langle S_{\rm max}\rangle$ and $\Delta_{\rm max}-\langle \Delta_{\rm max}\rangle$, 
respectively. Since $\langle S_{\rm max}\rangle$ and $\langle \Delta_{\rm max}\rangle$ are also
masses, this complication has no consequences, except that fitting procedures become less 
straight forward.\\
(iii) The critical exponents in Eqs.~(11-13) of \cite{Fan} look much more complicated. But 
although the notation ``$\beta_2$" in Eq.(11) suggests a relation with the order parameter $\beta$,
actually $\beta_2 = d-d_f$. The reason why we have $-d_f$ in our Eq.(3) instead of $d-d_f$ is 
that we defined $\Delta_n = S_n-S_{n-1}$, while $\Delta_n = (S_n-S_{n-1})/S_n$ in \cite{Fan}.\\
(iv) The subscript `2' on all critical exponents in Eqs.~(11-13) of \cite{Fan} should 
indicate that variances of observables for finite systems and shifts of these observables from
their infinite system values might follow different power laws, if standard FSS does not hold --
as was indeed found in \cite{Christensen} for correlation lengths in explosive percolation.
As shown in \cite{Fan,Li2023}, this is still true if one uses an 
event-based FSS analysis as in the present paper (whence the title of \cite{Li2023}
is somewhat misleading). Thus Eq.~(\ref{P_p}) must be changed for explosive percolation to allow
for two different exponents differents instead of the single $\nu$.

\section{Results}

\subsection{Discontinuous Transition}

We first show results for the discontinuous percolation cluster aggregation model (DCA) of \cite{Cho}. 
In this model, one starts with $N$ nodes and zero links. Links are inserted in 2 steps:\\
(i) choose two clusters, with probabilities $q_i \propto s_i^\omega$ ($i=1,2$), where $s_i$ is the 
mass of cluster $i$ and $\omega\in[0,1]$ is a constant;\\
(ii) on each cluster choose a random point, and link these two points, if they are not yet linked. If
they are already linked, choose another pair and repeat.\\
If the two chosen clusters are the same, this has no effect on the cluster distribution. If they are 
different, however, they are joined by the new link.\\
This is repeated until all clusters are joined into one big cluster.

Notice that this would be exactly the cluster growth in the ER model, if $\omega=1$. Interesting new
results are found for $\omega<1$. We and Fan {\it et al.} \cite{Fan} simulated only the 
extreme case $\omega=0$. In this case the algorithm can be simplified drastically -- in particular, 
if the NZ algorithm is used: In each NZ step, two different clusters are chosen randomly, and joined by 
adding a pointer from the root of one of them to the root of the other. Notice, however, that this
simplification -- which does not change $P_S$ and $P_\Delta$ -- changes the definition of the control
parameter, as no links are inserted which do not join two clusters.

It turns out that this leads in many events to a percolation transition (largest jump) when the last 
two clusters get joined, and since each of these two clusters has in general $O(N)$ sites, the 
transition is discontinuous. Compared to discontinuous transitions in models with cooperativity 
\cite{Dodds,Janssen,Dorogovtsev2006,Bizhani,Cai} or with dependency links \cite{Parshani,Son},
where the transitions usually happen at $p_c < 1$, this transition is a bit trivial (like the 
OP transition in $d=1$ at $p_c = 1$), but it is formally a genuine discontinuous transition.

\begin{figure}
  \begin{center}
  \psfig{file=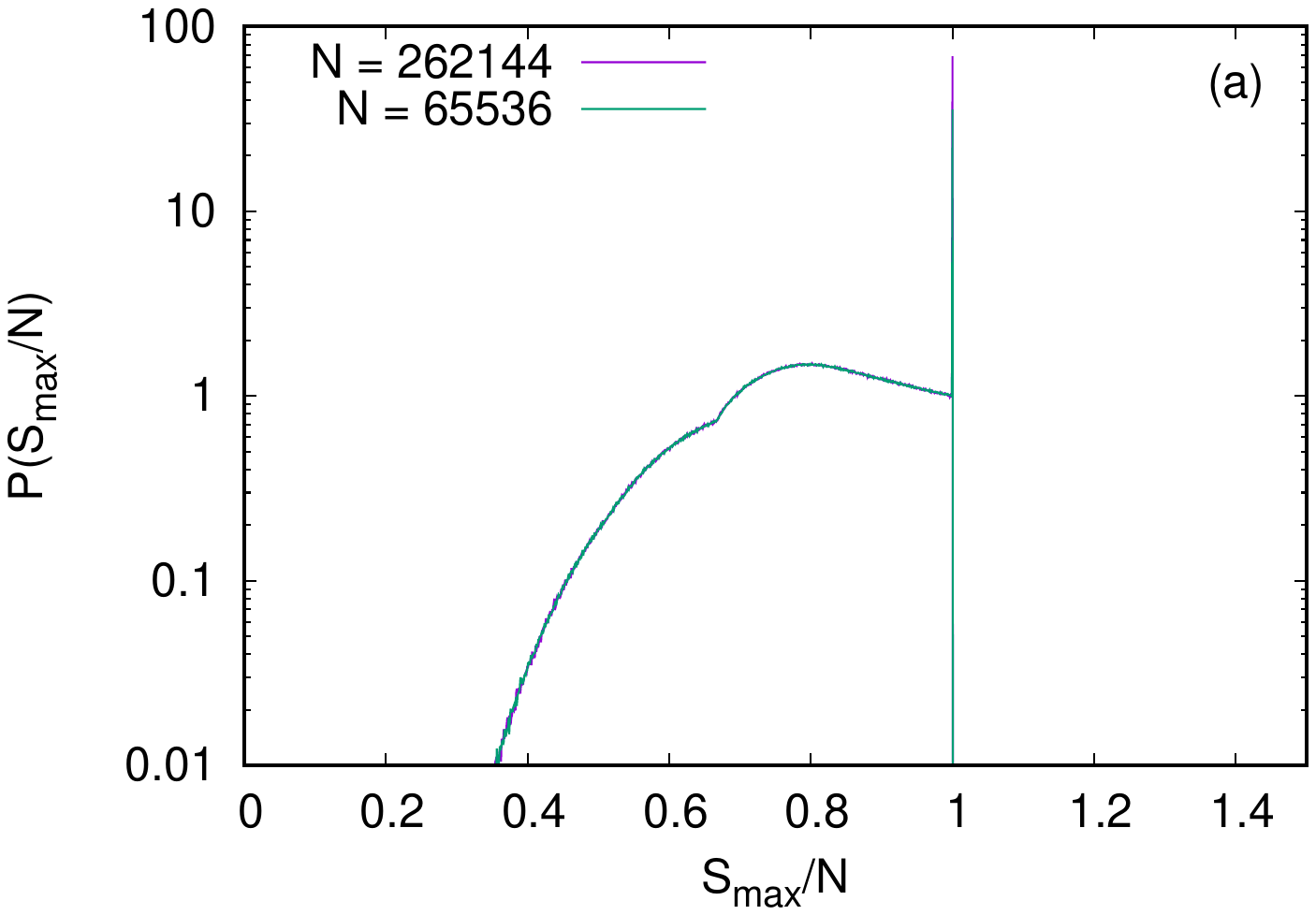,width=6.4cm, angle=0}
  \vglue -17pt
  \psfig{file=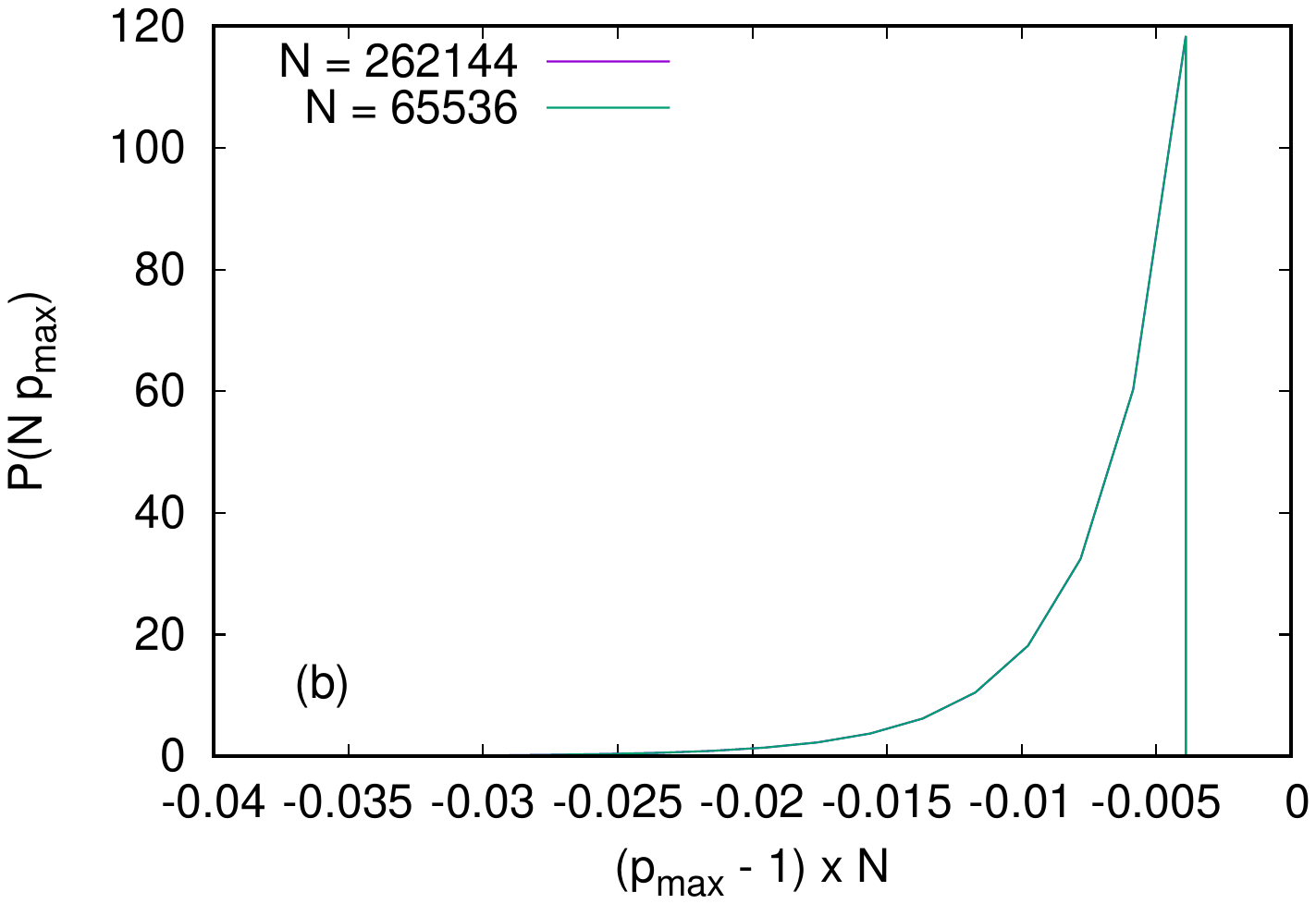,width=6.4cm, angle=0}
  \vglue -17pt
  \psfig{file=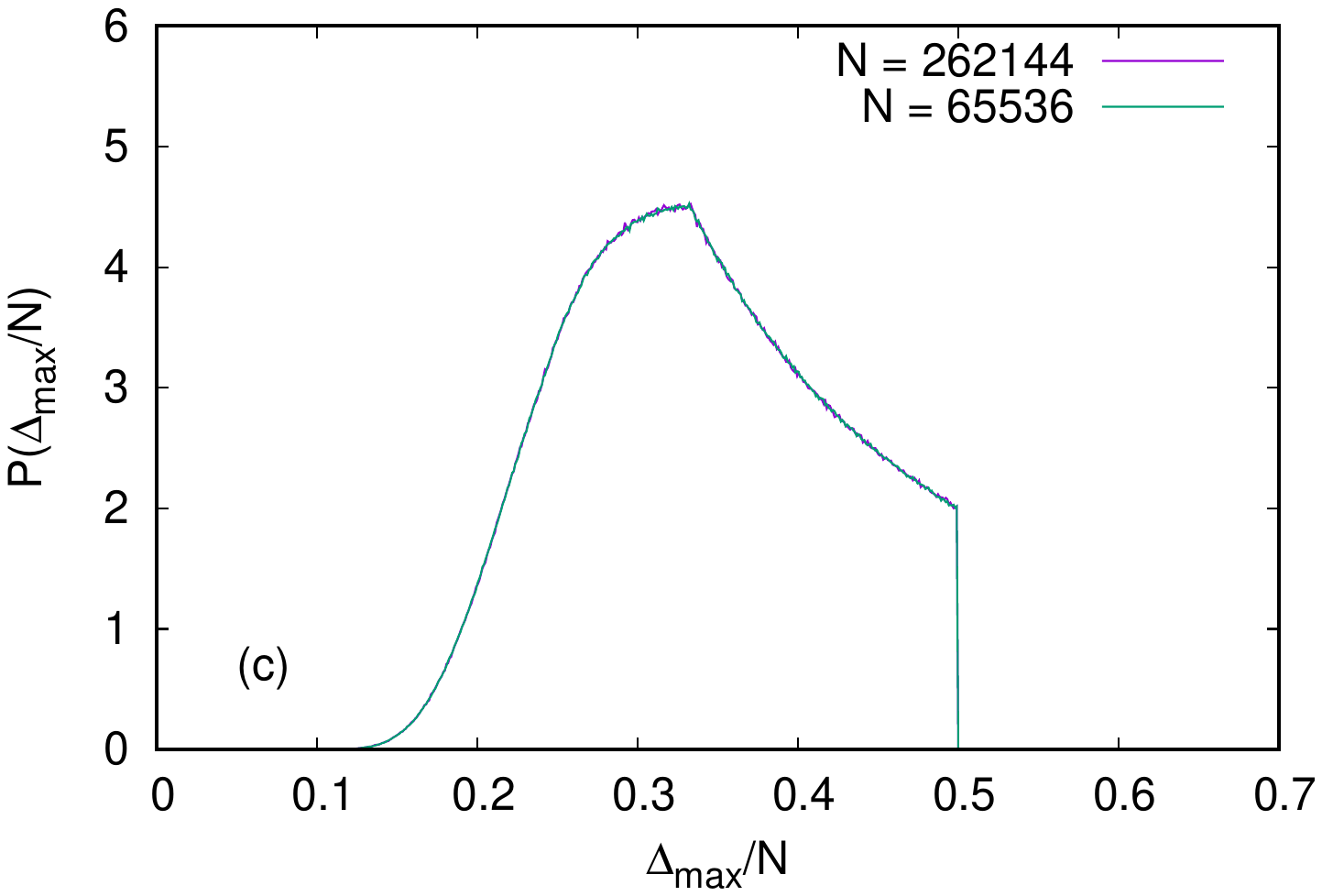,width=6.4cm, angle=0}
  \vglue -17pt
  \caption{Distributions of $S_{\rm max}/N$ (panel a), of $p_{\rm max}N$ (panel b), and of
	  $\Delta_{\rm max}/N$ (panel c) for the DCA model of \cite{Cho} at $\omega=0$. Because we 
	  used an optimized algorithm where bonds are inserted only when needed to join two clusters,
	  $P_p(p_{\rm max})$ differs from the distribution shown in \cite{Fan}, but the other two 
	  distributions should be the same.}
    \label{DCA}
  \end{center}
  \vglue -7pt
\end{figure}

Distributions $P_S(S_{\rm max}), P_p(p_{\rm max})$, and $P_{_\Delta}(\Delta_{\rm max})$ for $N=2^{16}$ 
and $N=2^{18}$
are shown in Fig.1. The mass distribution $P_S(S_{\rm max})$ shows a delta-peak at $S_{\rm max}$ due
to the fact that there is a finite chance that the maximal jump occurs exactly at $p=1$. Accordingly,
$P_p(p_{\rm max})$ and $P_{_\Delta}(\Delta_{\rm max})$ show jumps at $p_{\rm max}=1$ and at 
$\Delta_{\rm max}=1/2$. The latter is maybe the most obvious indication that the claim of \cite{Fan}
that this is a Gumbel distribution (which has an exponential right hand-tail) is wrong. It is not
clear why Fan {\it et al.} made this claim, since our Fig.1c agrees perfectly with their Fig.4a.
The distribution of $S_{\rm max}$ shown in Fig.4c of \cite{Fan} lacks the delta-peak, but shows the 
same sharp cut-off as our Fig.~1a. Also from this, it should be clear that the universality claim 
of Fan {\it et al.} is untenable. Finally, our distribution of $p_{\rm max}$ differs markedly from 
that of Fan {\it et al.}, but as we pointed out above, this was to be expected.

\begin{figure}
  \begin{center}
  \psfig{file=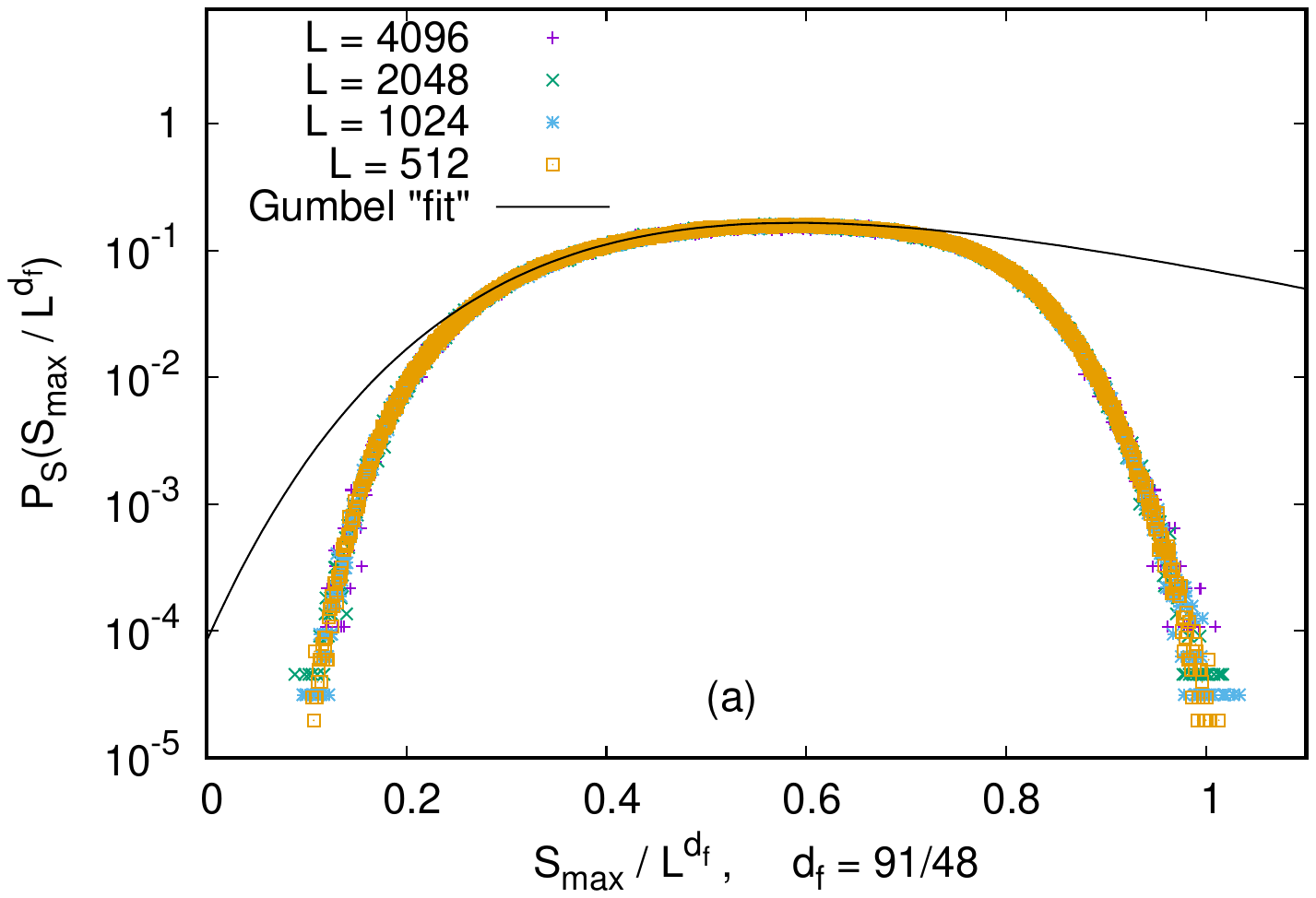,width=6.4cm, angle=0}
  \vglue -17pt
  \psfig{file=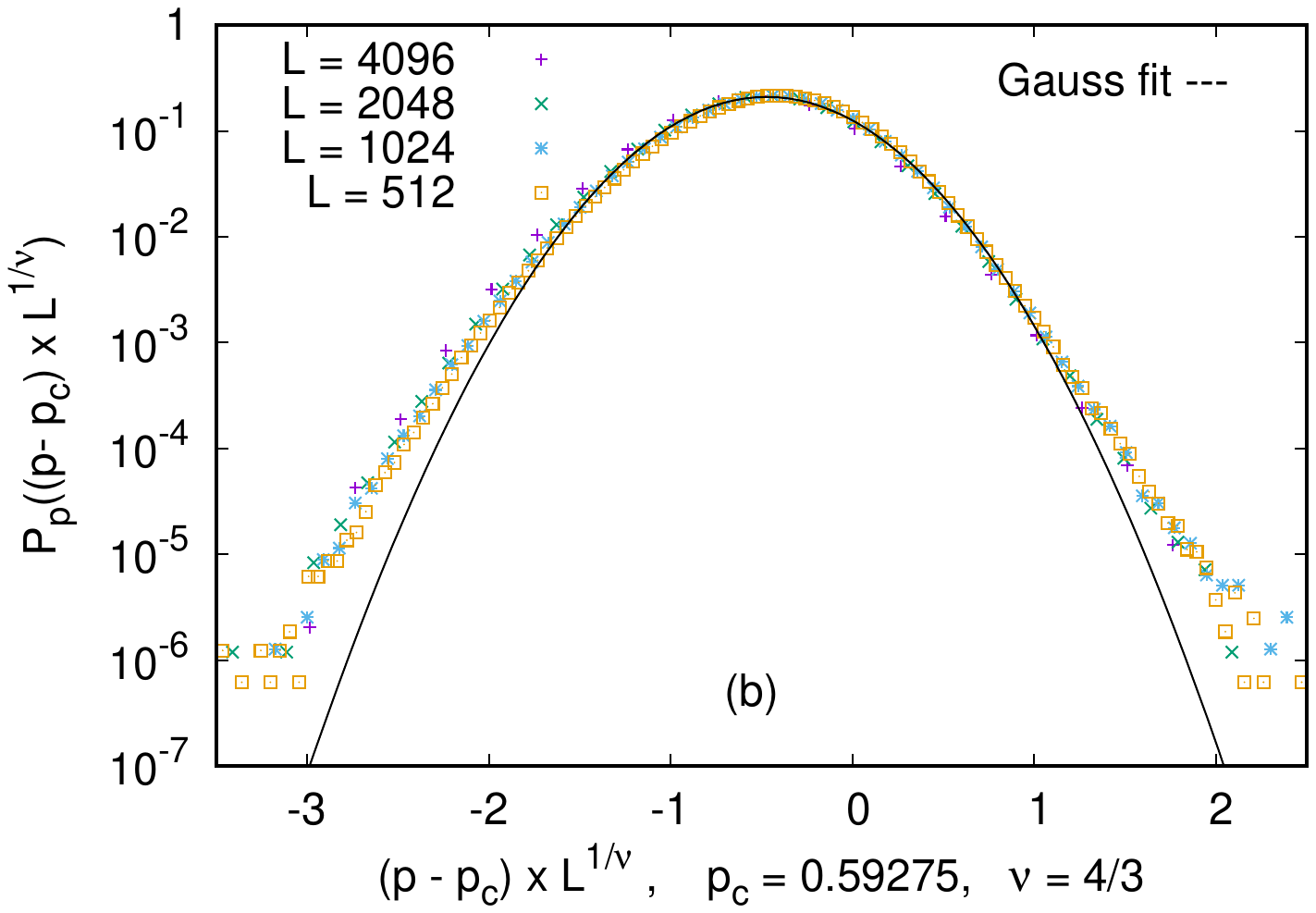,width=6.4cm, angle=0}
  \vglue -17pt
  \psfig{file=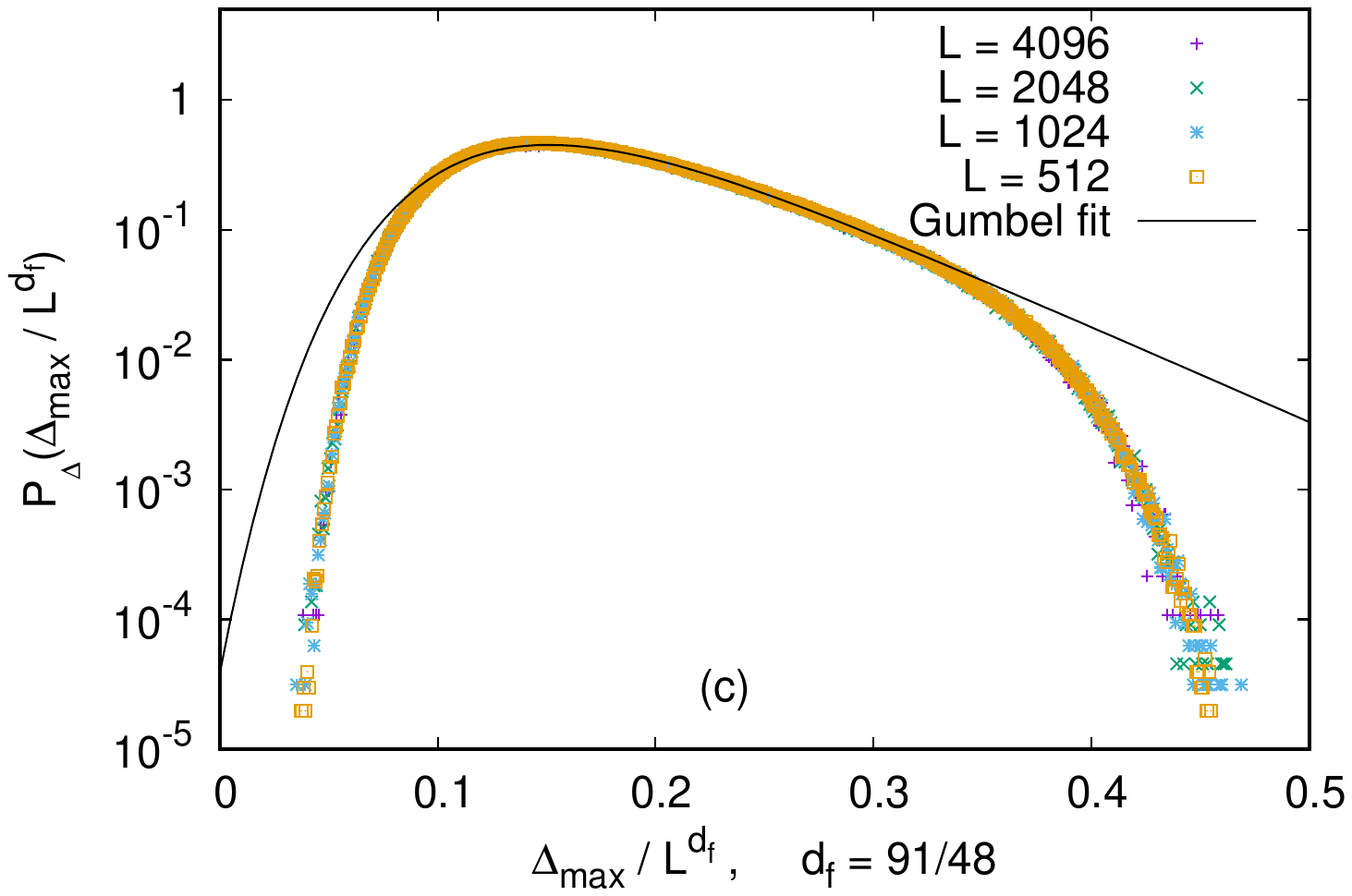,width=6.4cm, angle=0}
  \vglue -17pt
  \caption{Distributions of $S_{\rm max}/N$ (panel a), of $p_{\rm max*N}$ (panel b), and of
          $\Delta_{\rm max}/N$ (panel c) for ordinary site percolation in $d=2$. In contrast to Fig.~1,
	  normalization is now arbitrary, and log-scales are used on the y-axis in all panels. In panels
	  (b) and (c) we draw also best-fit Gauss (panel (b)) and Gumbel (panel (c)) distributions.
	  Because we could not produce even a modestly acceptable Gumbel fit for panel (a), we plot there
	  the best-fitting Gumbel distribution with the $\beta$-parameter obtained in \cite{Fan} for 
	  2-d bond percolation (due to universality, site and bond percolation should have identical
	  scaling functions).}
    \label{2-d_site}
  \end{center}
  \vglue -7pt
\end{figure}

Since the transition in the DCA model is not hybrid, the critical exponents are not anomalous.
We see clearly $\langle S_{\rm max}\rangle \sim \langle \Delta_{\rm max}\rangle \sim N$ in agreement 
with Fan {\it et al.}, and $1-\langle p_{\rm max}\rangle \sim 1/N$. The latter disagrees with 
the claim of Fan {\it et al.} that the model shows two different correlation length exponents, and 
might be due to our optimized algorithm.

\subsection{Ordinary Site Percolation in $d=2$}

In the previous subsection we saw that the claim of Fan {\it et al.} is wrong in a case where the 
distributions $P_S(S_{\rm max})$ and $P_{_\Delta}(\Delta_{\rm max})$ could not be Gumbel simply by the 
fact that they were sharply cut off at finite arguments, while the Gumbel distribution has an exponential
tail. This could have been a simple oversight. 

The next case, OP in $d=2$, is more serious. In Fig.2 we plot $P_S(S_{\rm max}), P_p(p_{\rm max})$, 
and $P_{_\Delta}(\Delta_{\rm max})$ for $N=2^{16}, 2^{18}, 2^{20}$, and $2^{22}$. To obtain a data
collapse, we used on the $x$-axes the variables $S_{\rm max}/L^{d_f}, (p_{\rm max}-p_c)L^{1/\nu}$,
and $\Delta_{\rm max}/L^{d_f}$ with known values of $d_f,\nu$, and $p_c$. Normalization is arbitrary.

\begin{figure}[h]
  \begin{center}
  \psfig{file=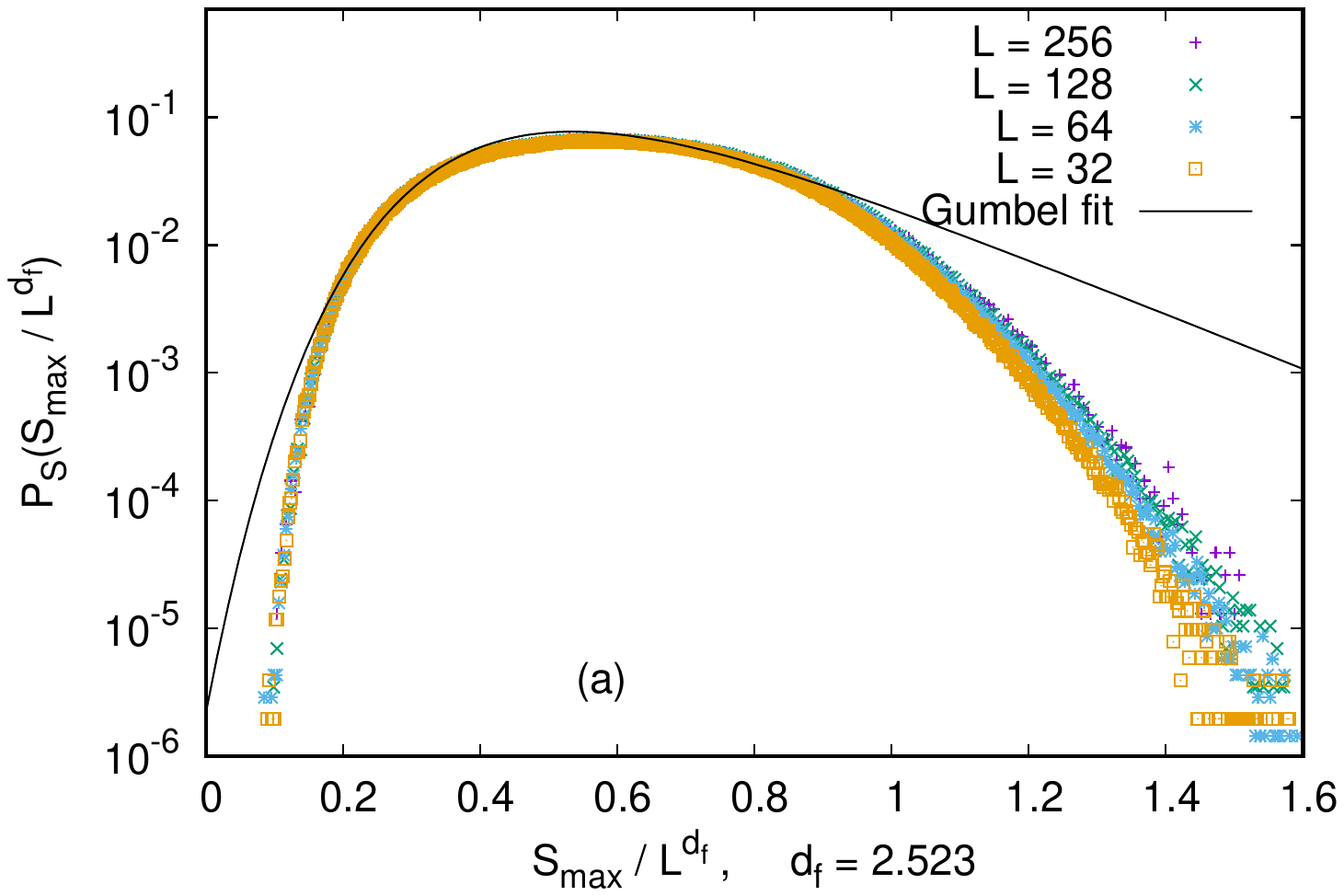,width=6.4cm, angle=0}
  \vglue -17pt
  \psfig{file=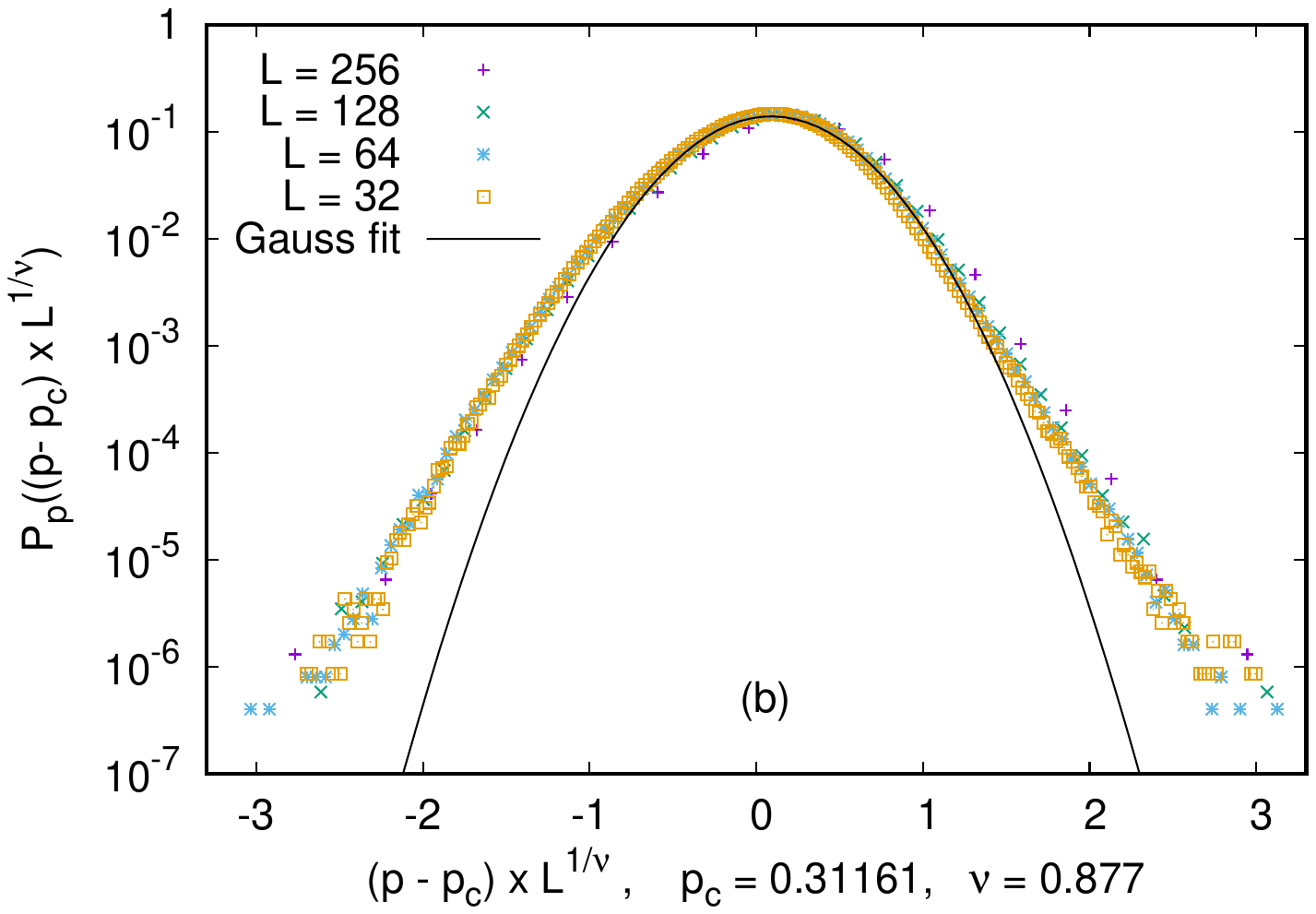,width=6.4cm, angle=0}
  \vglue -17pt
  \psfig{file=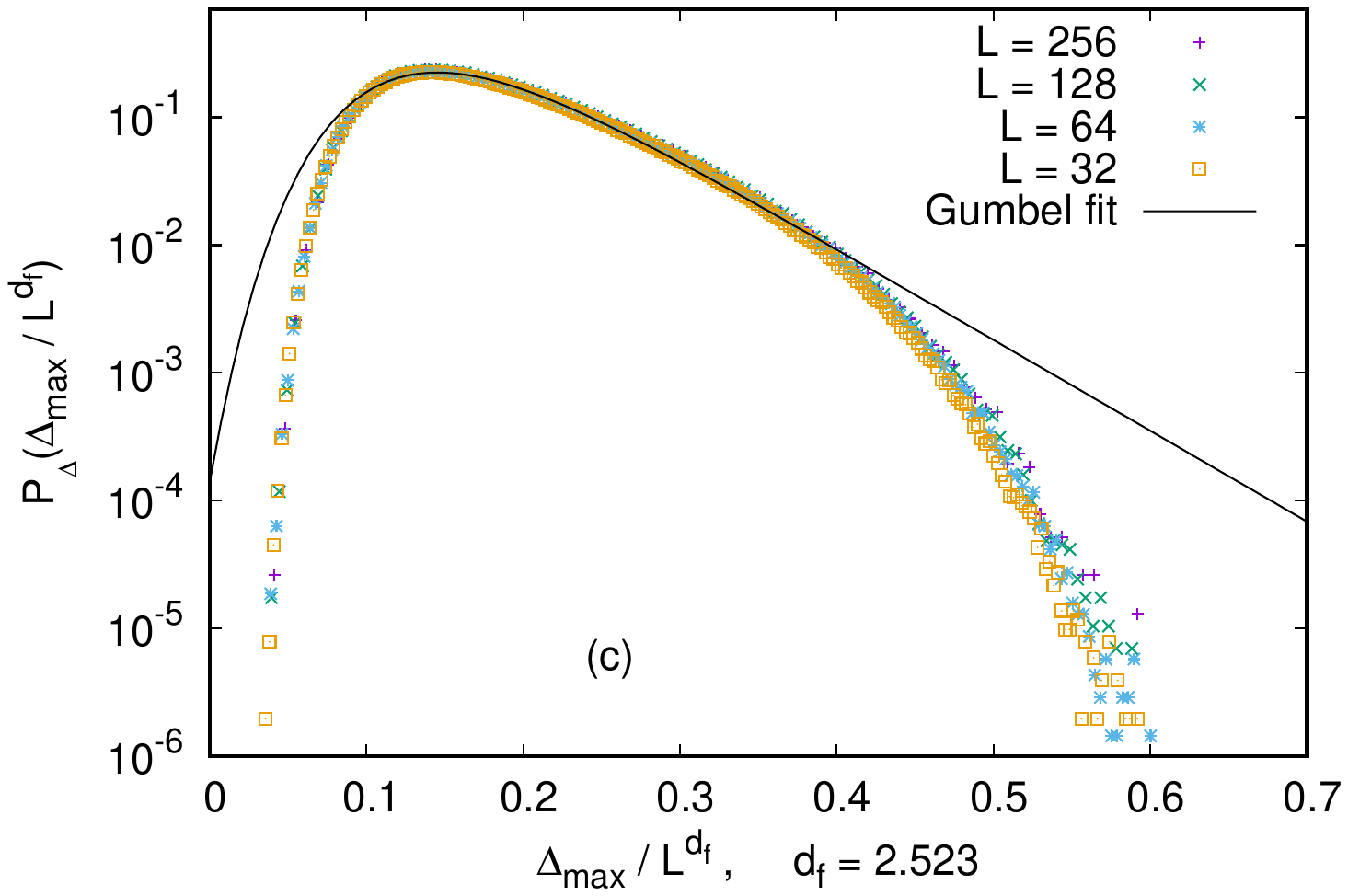,width=6.4cm, angle=0}
  \vglue -17pt
  \caption{Distributions of $S_{\rm max}/N$ (panel a), of $p_{\rm max*N}$ (panel b), and of
          $\Delta_{\rm max}/N$ (panel c) for ordinary site percolation in $d=3$. Normalization is 
	  arbitrary as in Fig.~2, and log-scales are again used on the y-axis in all panels. In all 
	  panels, the continuous curves represent best eyeball fits to the data, in which the tails of 
	  the distributions are neglected.}
    \label{3-d_site}
  \end{center}
  \vglue -7pt
\end{figure}

Our data in Fig.2 agree perfectly with the data for 2-d bond percolation shown in Fig.~2 of \cite{Fan}.
But in none of the panels, the fits with Gumbel resp. Gauss distributions are acceptable. For panel
(b) the disagreement is least severe, and it would have gone unnoticed if we had not used a log scale
on the y-axis. The disagreement with a Gumbel fit is much worse for panel (c), but for panel (a) it 
is so large that it is unclear how this could have gone unnoticed in \cite{Fan}. Notice that the 
Gumbel $\beta$-parameter obtained in \cite{Fan} (called there $\omega_2$, see \cite{suppl}) 
is indeed the one which gives the best fit in the left half of the plot, but it produces
nonsense on the right half. Nevertheless, in Fig.~2f of \cite{Fan} a supposed Gumbel distribution 
is drawn which gives a nearly perfect fit.

\begin{figure}[h]
  \begin{center}
  \psfig{file=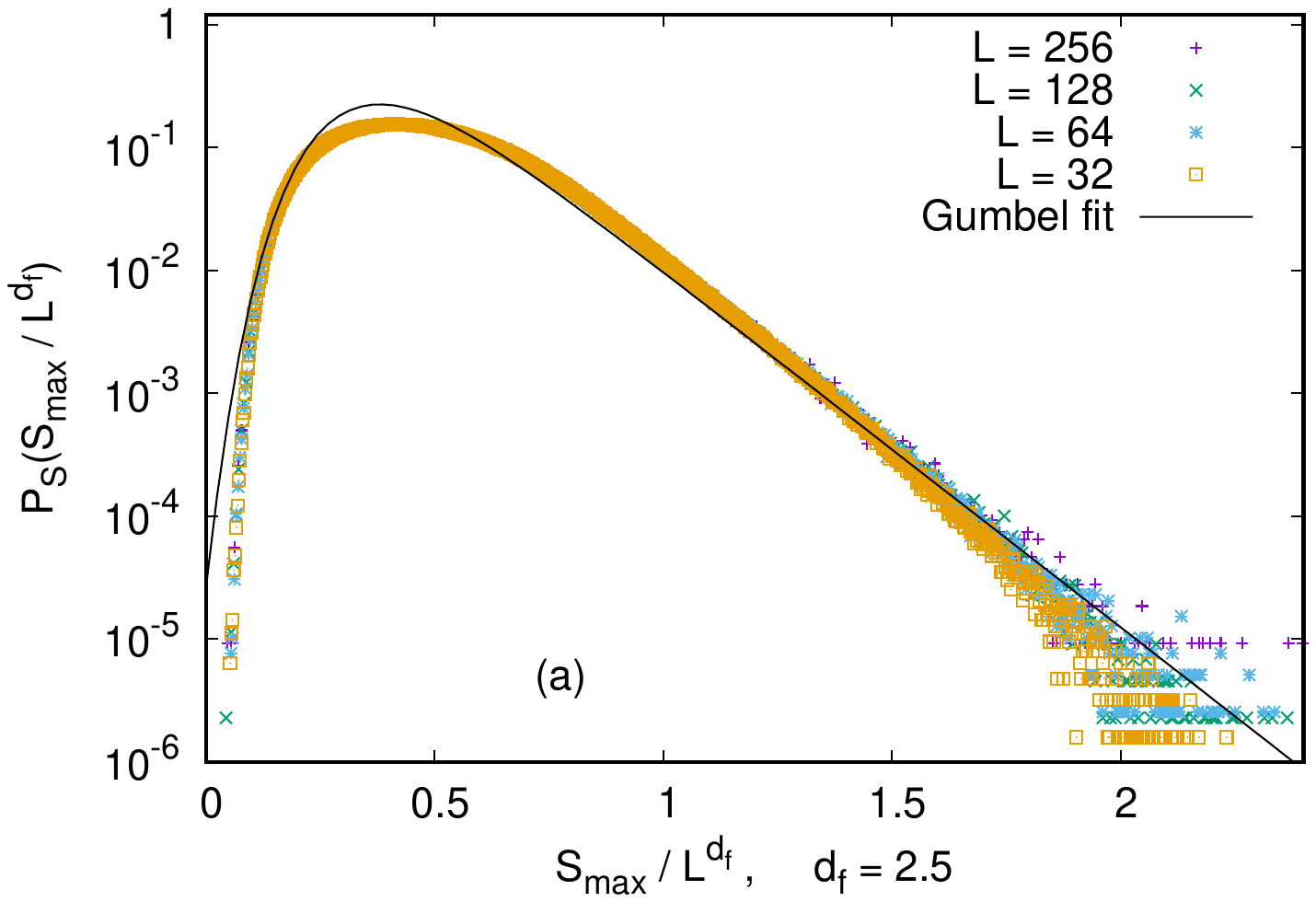,width=6.4cm, angle=0}
  \vglue -17pt
  \psfig{file=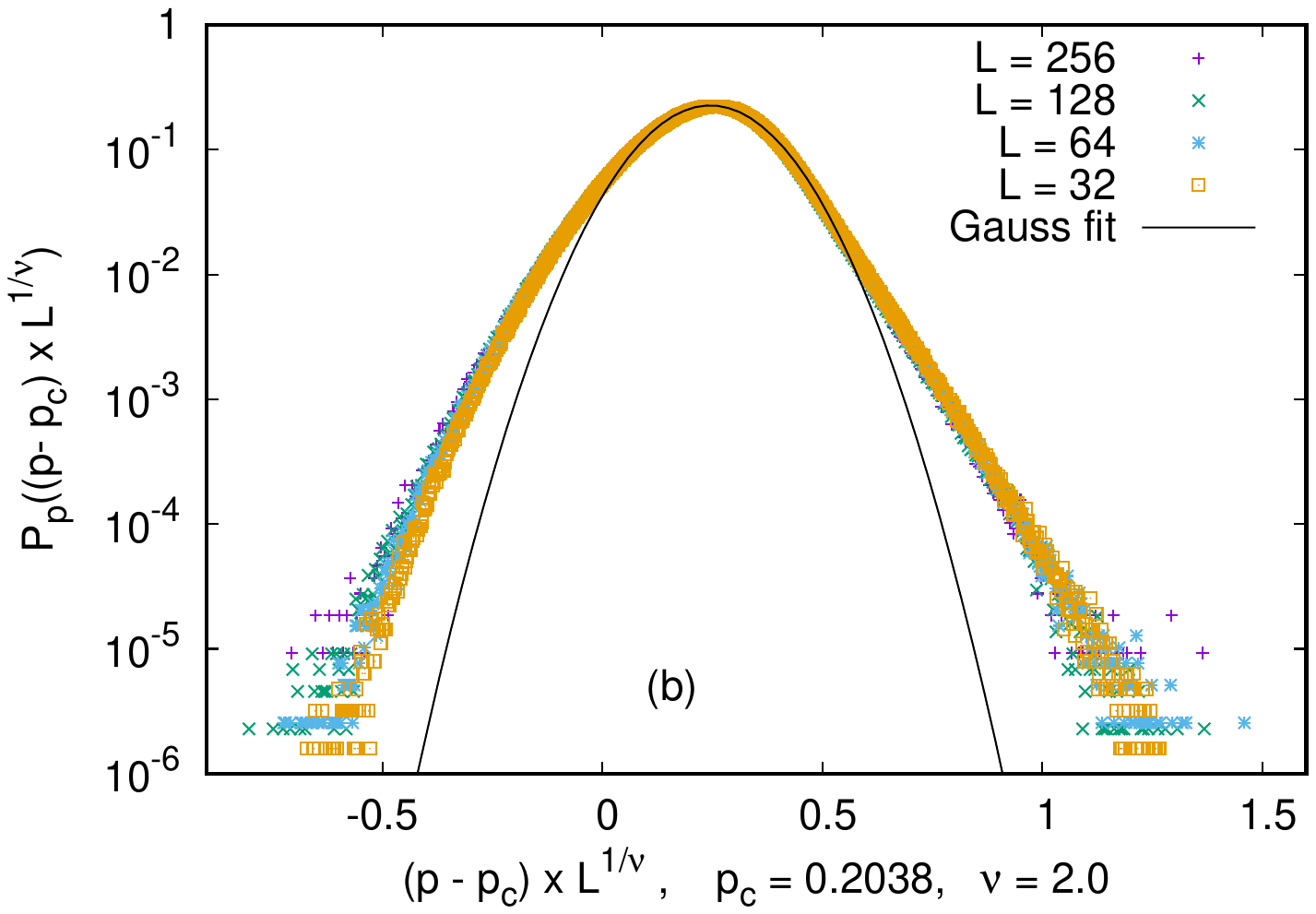,width=6.4cm, angle=0}
  \vglue -17pt
  \psfig{file=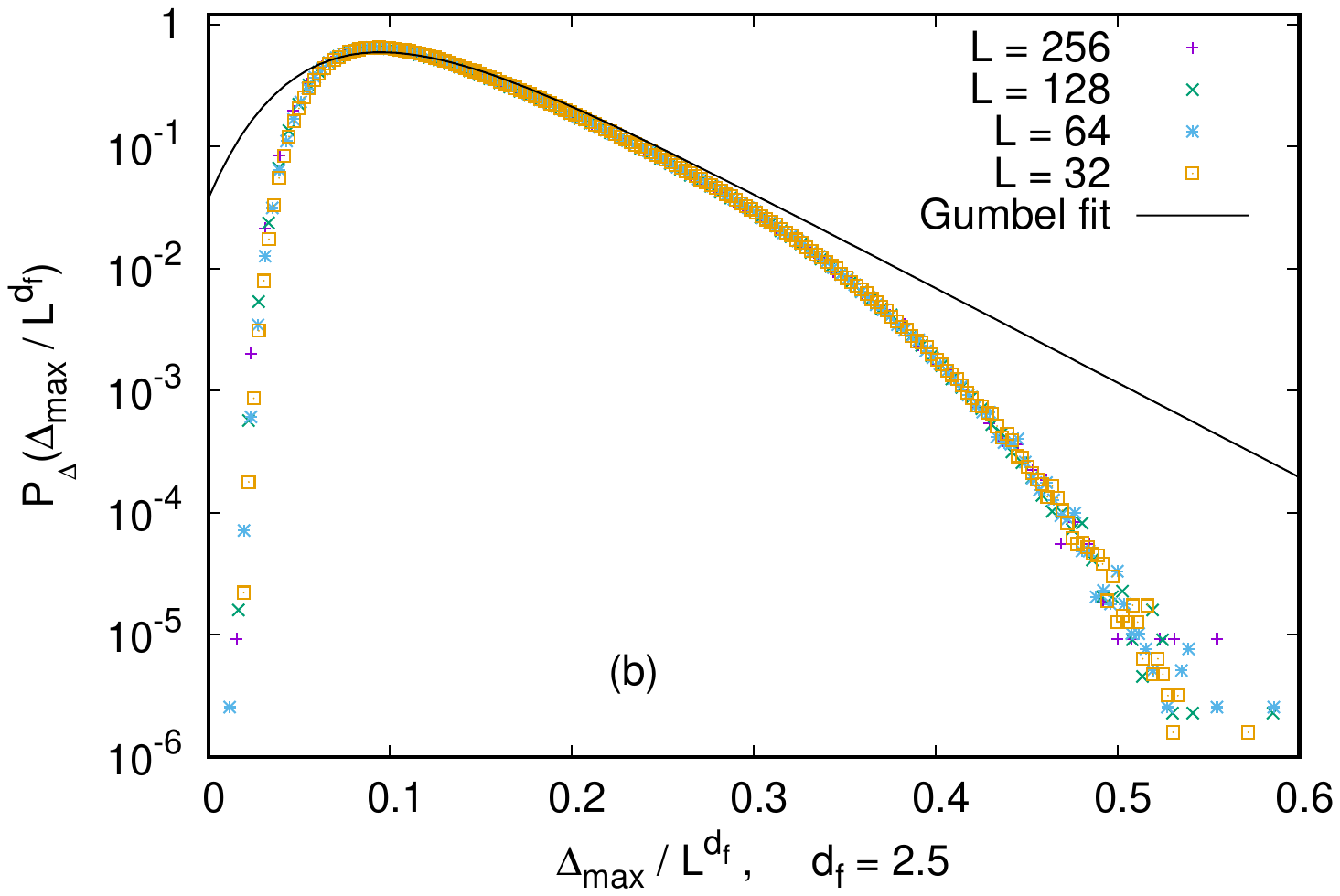,width=6.4cm, angle=0}
  \vglue -17pt
  \caption{The same three distributions for aftermath percolation in $d=3$. Normalization is
          arbitrary as in the previous figures, and log-scales are again used on the y-axis in all panels. 
	  In panel (a) one sees a clear exponential tail, therefore the (eyeball) Gumbel fit is made so that 
	  this tail gets a substantial weight.}
    \label{3-d_aftermath}
  \end{center}
  \vglue -7pt
\end{figure}

\subsection{Three-Dimensional Lattice Models}

The next models also show that the claim of Fan {\it et al.} is untenable, but the deviations from 
extreme value distributions are less pronounced than in the previous models. Data for site percolation
in $d=3$ are shown in Fig.~3. Again, the raw data agree fairly well with those of \cite{Fan} (see 
Fig.~S2 in \cite{suppl}), but the fits to Gauss and Gumbel distributions are definitely 
unacceptable. Finally, we should point out that the right-hand tails of $P_S$ and $P_{_\Delta}$ 
slightly rise with increasing $L$. But this rise seems too slow to cure the problem for large $L$, 
and there is no similar rise of the left-hand tails.

The second $3-d$ lattice universality class studied here is aftermath percolation, introduced in 
\cite{aftermath}. It is in the universality class of pacman percolation \cite{Abete}, but is easier
to simulate with the NZ algorithm. The critical exponents were conjectured in \cite{aftermath} to be 
exactly $\beta=1, \nu=2,$ and $d_f = 5/2$. The three distributions for this universality class are 
shown in Fig.~4. This time, $P_S$ shows a rather clear exponential tail, and can thus be fitted 
fairly well by a Gumbel distribution (although such a fit is statistically unacceptable). On the 
other hand, $P_{_\Delta}$ is very poorly represented by a Gumbel distribution.

\begin{figure}[h]
  \begin{center}
  \psfig{file=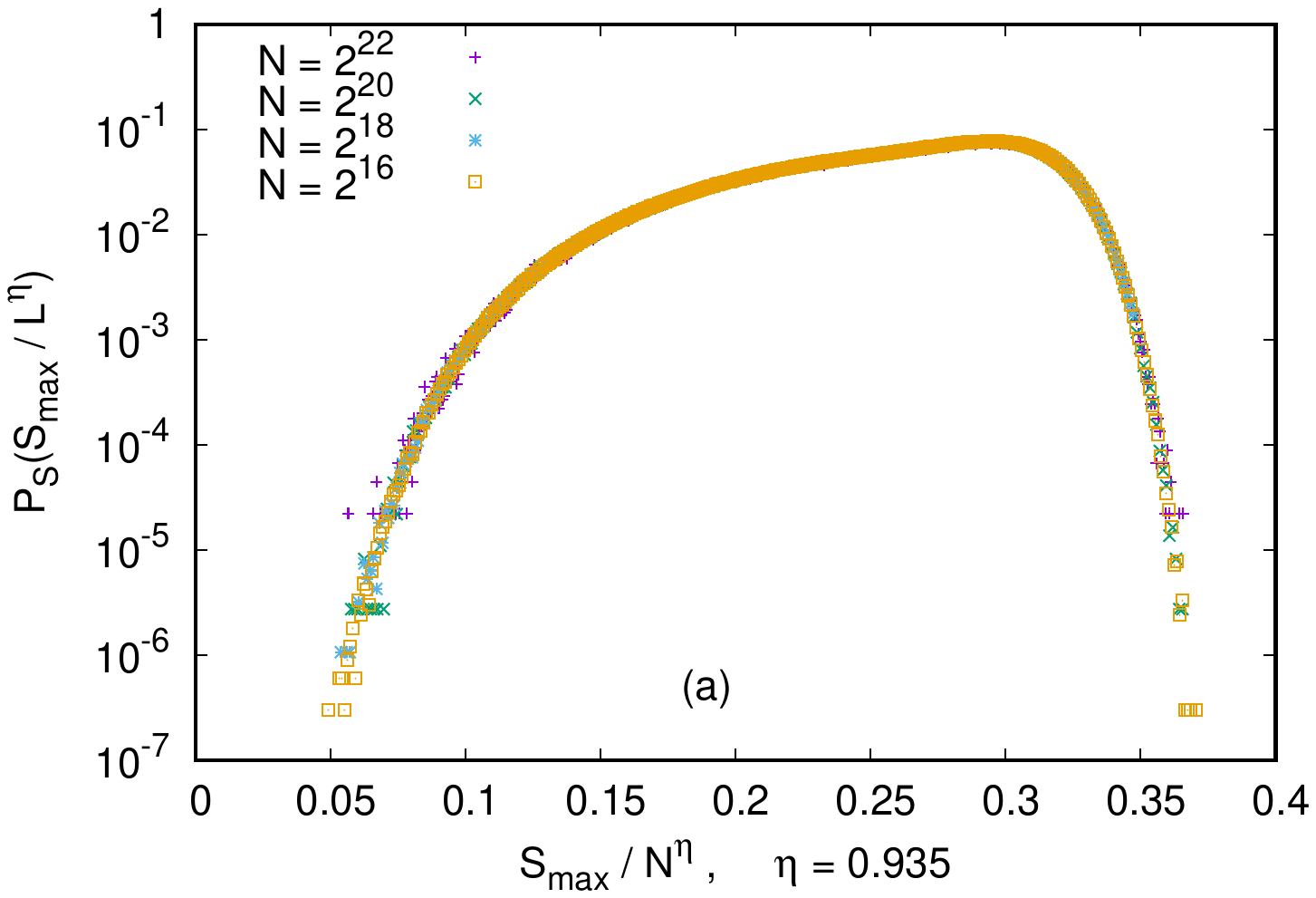,width=6.4cm, angle=0}
  \vglue -17pt
  \psfig{file=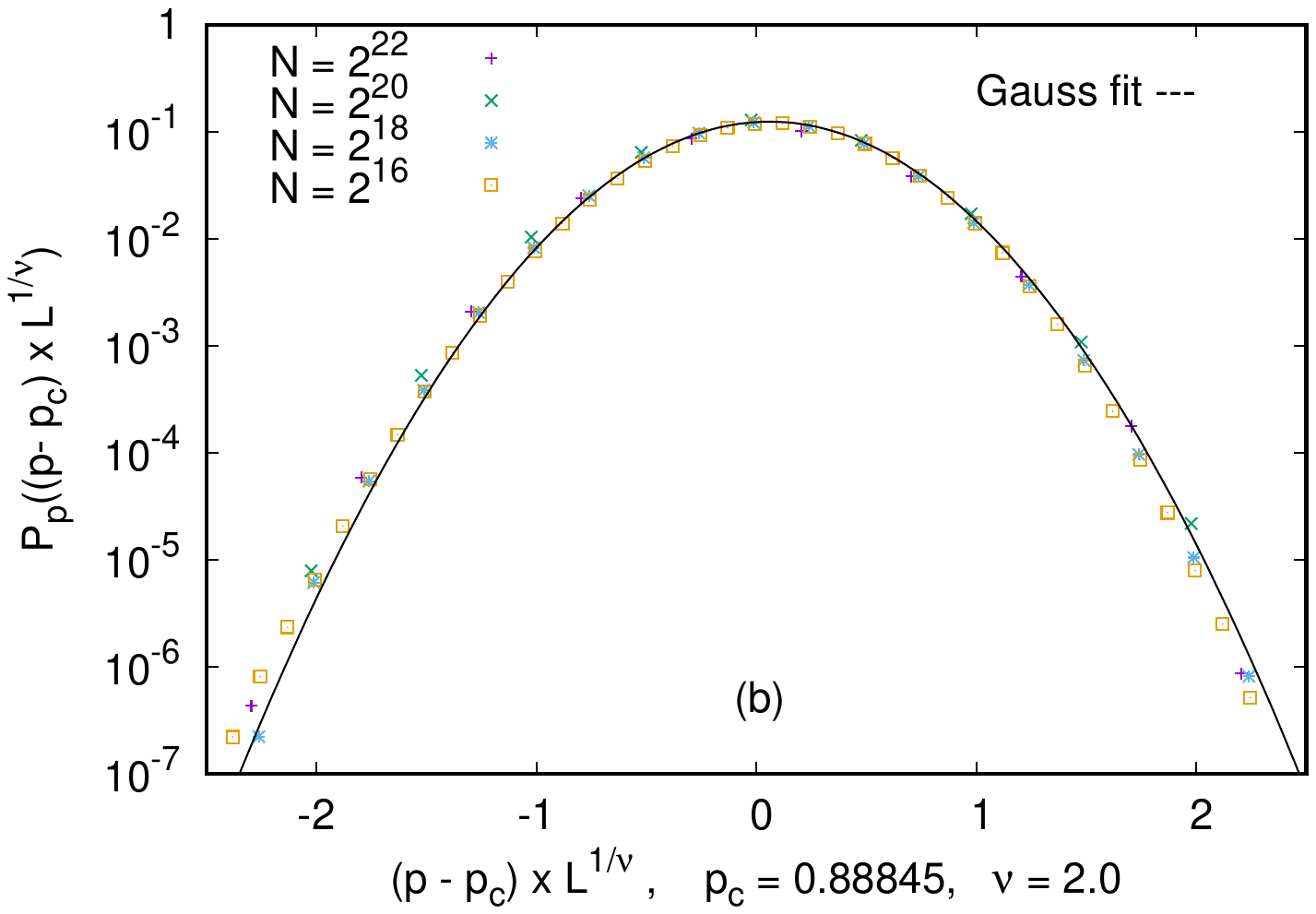,width=6.4cm, angle=0}
  \vglue -17pt
  \psfig{file=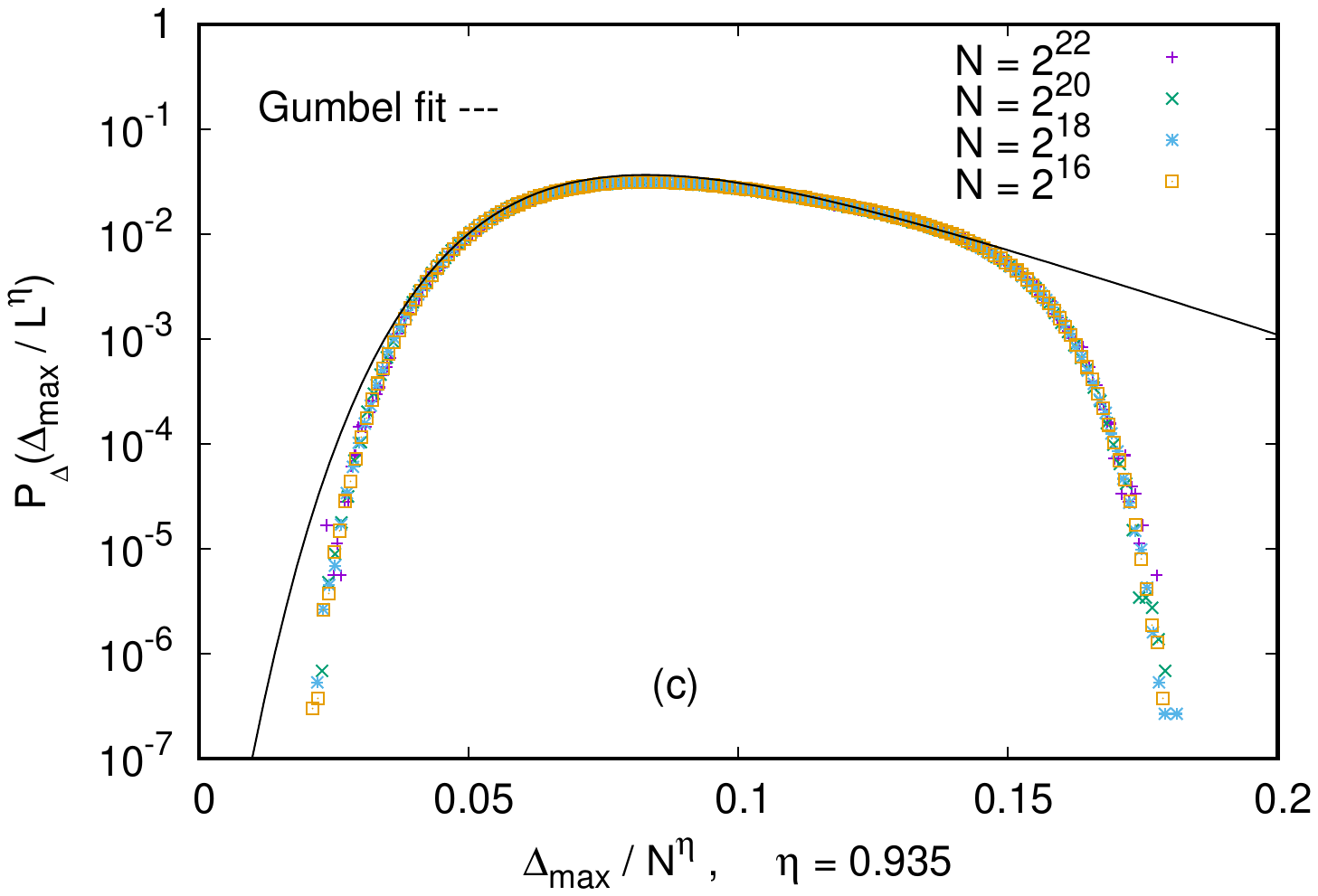,width=6.4cm, angle=0}
  \vglue -17pt
  \caption{The same three distributions for explosive percolation with the product rule. Normalization is
          arbitrary as in the previous figures, and log-scales are again used on the y-axis in all panels.
	  $P_S$ (shown in panel (a)) is so far from any Gumbel distribution that we refrained from showing
	  a fit, the fits in (b) and (c) are as before eyeball fits.}
    \label{explosive}
  \end{center}
  \vglue -7pt
\end{figure}

\subsection{Explosive Percolation}

There are a large number of explosive percolation models, which all use as a core ingredient the 
Achlioptas trick \cite{Achlioptas} to delay the formation of large clusters. The critical scaling of 
four of them was studied in detail in \cite{Christensen}, where it was found that they all had vastly 
different order parameter exponents. Thus they belonged to different universality classes -- although 
$\nu$ (called $\Theta$ in \cite{Christensen}) was the same in the three off-lattice models studied in 
\cite{Christensen}. 

\begin{figure}[h]
  \begin{center}
  \psfig{file=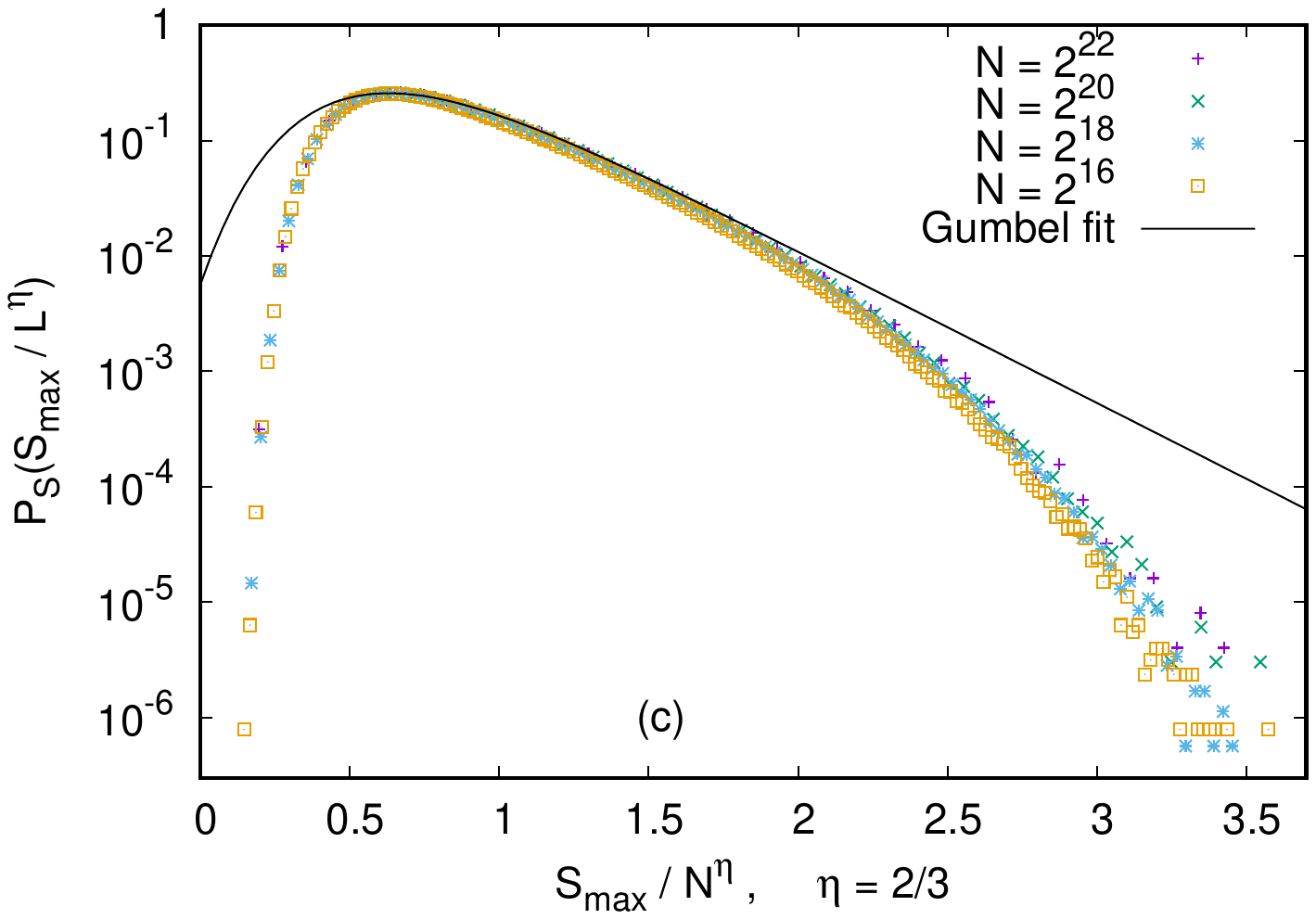,width=6.4cm, angle=0}
  \vglue -17pt
  \psfig{file=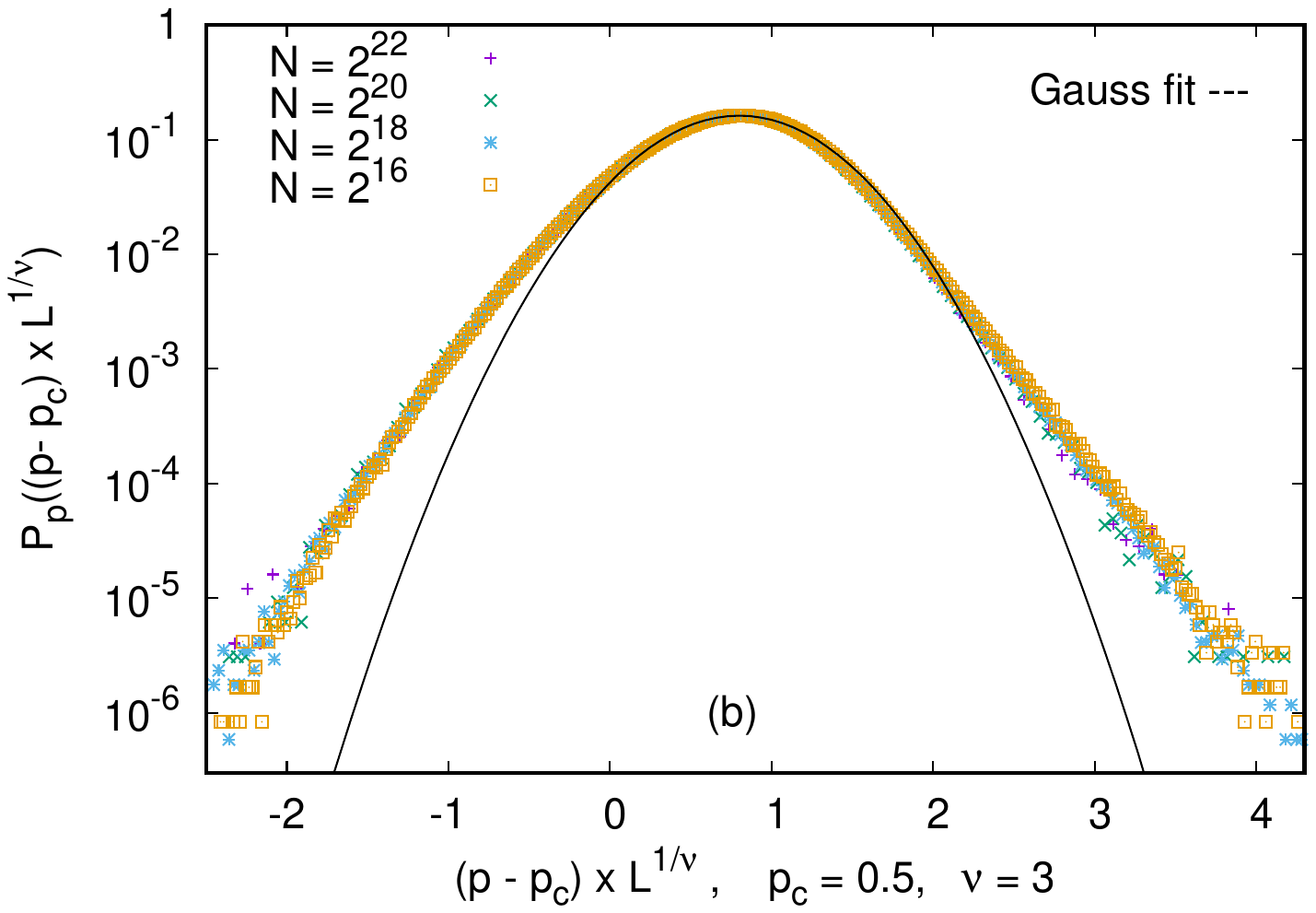,width=6.4cm, angle=0}
  \vglue -17pt
  \psfig{file=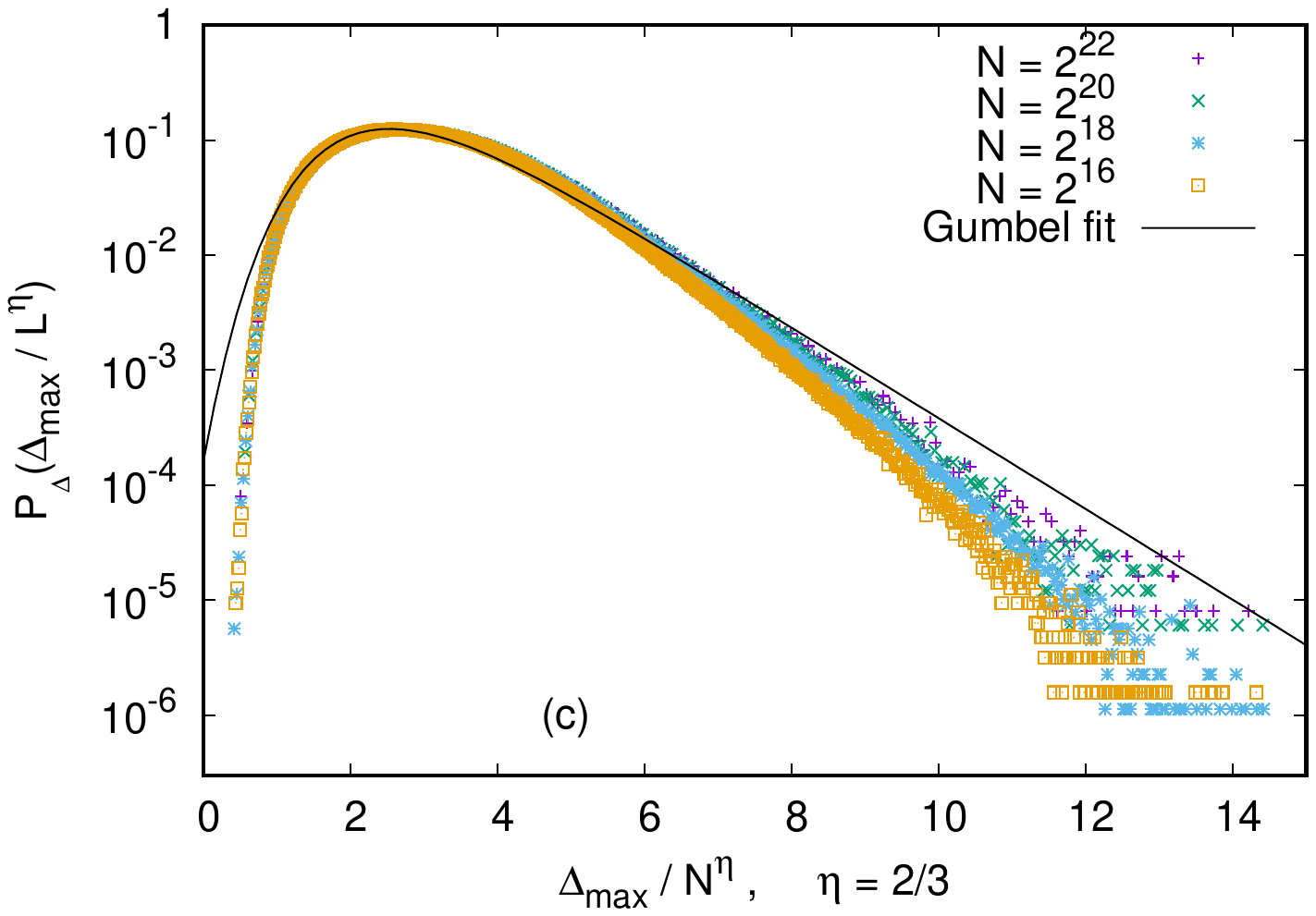,width=6.4cm, angle=0}
  \vglue -17pt
  \caption{The same three distributions for critical ER networks. Normalization is
          arbitrary as in the previous figures, and log-scales are again used on the y-axis in all panels.}
    \label{ER}
  \end{center}
  \vglue -7pt
\end{figure}

In the following we shall only show results for the ``product rule", the 
original model proposed in \cite{Achlioptas}. The three distributions for this model are shown in
Fig.~5. We see that $P_S$ is so far from a Gumbel that we didn't try to fit one. On the other hand,
$P_p$ is perfectly compatible with a Gaussian, as was conjectured in \cite{Li2023} \cite{foot1}. 
Also, this Gaussian is nearly perfectly centered at $p-p_c=0$, i.e. statistical fluctuations of the 
effective transition point $p_c(N)$ at large but finite $N$ are much bigger than its deviation from 
$p_c$, and scale with different exponents \cite{Christensen,Fan,Li2023}. This is in sharp contrast to 
what is found in most other second-order phase transitions, where both scale with the same exponent, 
$\sigma(p_c) \sim p_c(N)-p_c \sim N^{-1/\nu}$.

\subsection{Erd\"os-Renyi Networks}

Finally, we show the three distributions also for critical ER-networks. There the critical point is
at $p_c = 1/2$, where $p$ is the number of links divided by $N$, $\nu = 3$, and the cluster masses 
scale as $N^{2/3}$. All these are clearly seen in Fig.~6. Again we see that $P_p$ is not Gaussion,
and $P_S$ is not Gumbel. The situation is different for $P_{_\Delta}$. There, we see a strong finite 
size dependence in the right hand tail, so that this tail could well be fitted by a Gumbel for 
$N\to\infty$. On the other hand, there is no such $N-$ dependence in the left hand side of 
$P_{_\Delta}$, so that a Gumbel can be again excluded.

\section{Discussion}

We have performed extensive simulations of six different universality classes of (generalized)
percolation. It had been claimed in \cite{Fan} that certain distributions should in all 
such universality classes follow extreme value distributions. More precisely, this claim was 
made for what is called the ``event-based ensemble" in \cite{Li2023}, and the distributions are 
those of the largest cluster mass at the (event-based) pseudo-critical point, the ``gap" (largest
jump in the giant cluster mass), and the control parameter at the pseudo-critical point. It was 
claimed in \cite{Fan} that the first two should be Gumbel distributions, and the third Gaussian.

This was claimed to be true even if the transition is discontinuous, which would establish a novel
instance of super-universality extending over continuous {\it and} discontinuous transitions. But 
we found that the claim was untenable in {\it all} eighteen distributions we looked at, except 
for one: In explosive percolation with the product rule, the pseudo-critical control parameter 
values showed a Gauss distribution, as had been conjectured in \cite{Li2023}.

Apart from this negative result, we obtained very precise universal distributions, which could be 
very useful in future studies. We also verified in all cases the known and/or conjectured values 
of critical exponents. In particular this holds for aftermath percolation \cite{aftermath}, which
is in the universality class of pacman percolation, and where $\beta =1$ seems to hold for all
dimensions between $d=3$ and $d=6$ \cite{Fesh-pacman}.

Acknowledgements: M.F. acknowledges the support from the research council of the Alzahra University.
P.G. thanks Youjin Deng for discussions about explosive percolation.


\begin{thebibliography}{99}
\bibitem{foot1} The slight asymmetry seen in panel (b) is probably an artefact due to binning.
\end{thebibliography}
\end{document}